\documentclass[aps,prb,twocolumn,superscriptaddress]{revtex4-1}
\usepackage{amsmath}
\usepackage{xcolor}
\usepackage{graphicx}
\usepackage{natbib}
\usepackage{hyperref}

\begin{document}

\title{Thermal Kosterlitz-Thouless transitions in the $1/r^2$ long-range ferromagnetic quantum Ising chain revisited}
\date{\today}
\author{Stephan Humeniuk}
\affiliation{Institute for Theoretical Physics III, University of Stuttgart, 70550 Stuttgart, Germany} 
\altaffiliation[Current address: ]{Beijing National Laboratory for Condensed Matter Physics and Institute of Physics, Chinese Academy of Sciences, Beijing 100190, China}

\begin{abstract}

For the inverse square long-range ferromagnetic Ising chain in a transverse field, the thermal
phase boundary of the floating Kosterlitz-Thouless phase is obtained for 
several values of the transverse field down to the quantum critical point. 
The sharp domain walls in the classical model are increasingly smeared out by 
the transverse field, which is evidenced by a pronounced broadening of the non-universal bump
in the specific heat. 
The discernability of KT critical scaling in finite-size simulations is discussed. 

\end{abstract}
\maketitle

\section{Introduction}

The one-dimensional (1D) ferromagnetic Ising chain with interactions decaying as $1/r^2$
is a cornerstone of statistical mechanics and the theory of phase transitions:
It deserves special attention from a historic point of view \cite{KosterlitzNobelLecture},
being the first physical system for which a Kosterlitz-Thouless (KT)
transition was described quantitatively 
and to which an early version of the renormalization group (RG) 
was applied \cite{Anderson1970,Kosterlitz1974}.
Furthermore, the 1D setting has allowed to study system sizes 
that span six orders of magnitude \cite{Luijten2001,Fukui2009}, which has facilitated 
precise verification of RG predictions and scaling laws. 

Here, the question is revisited \cite{Dutta2001, Fukui2009} whether quantum fluctuations change the nature 
of this thermal Kosterlitz-Thouless phase transition. 
While this might be answered in the negative on very general grounds \cite{Dutta2001}, 
I provide detailed quantum Monte Carlo calculations 
illustrating and corroborating what such a statement means as the strength
of the quantum fluctuations is increased. 

The Hamiltonian for general decay power $\alpha$ of the long-range interactions is
\begin{equation}
  \mathcal{H} = - \sum_{i < j=1}^{L} J_{ij} S_i^{z} S_j^{z} 
  - \Gamma \sum_{i=1}^{L} S_i^{x},
  \label{eq:LRTFI}
\end{equation}
where $(S_i^x, S_i^y, S_i^z)$ are spin-$\frac{1}{2}$ operators,
$J_{ij} = \frac{J}{|i-j|^{\alpha}}$ are long-range ferromagnetic ($J>0$) interactions 
and $\Gamma$ is the strength of the transverse field, which introduces quantum fluctuations. 
The phase transitions of the classical model ($\Gamma = 0$) are well-established. 
For $\alpha \le 2$, the model has a finite temperature phase transition 
from a paramagnet to a ferromagnet, while for $\alpha > 2$ 
there is no long-range order at any finite temperature \cite{Ruelle1968, Dyson1969}. 
The boundary case $\alpha = 2$ exhibits a Kosterlitz-Thouless 
transition \cite{Kosterlitz1974,Bhattacharjee1981,Froehlich1982,Luijten2001}
due to the presence of topological defects 
with a mutual interaction that depends logarithmically on their
distance. 
Based on general properties of thermal phase transitions 
in quantum systems it has been argued \cite{Dutta2001} 
that this picture is not quatlitatively altered by quantum fluctuations. 
The finite-temperature properties of the \emph{quantum} $1/r^2$ ferromagnet
were partially studied in Ref.~\cite{Sandvik2003} for $\Gamma=0.5$, and for the case of 
$\Gamma = 1$ on very large system sizes in Ref.~\cite{Fukui2009}, demonstrating that 
the KT transitions survive under weak quantum fluctuations. 
Note that an essentially equivalent problem setting arises for a quantum Ising chain coupled 
to a bosonic bath \cite{Werner2005}
where integrating out the bath degrees of freedom results in long-range interactions 
in imaginary time which decay asymptotically like an inverse square power law.

The present work completes 
the phase boundary in the full temperature-transverse field plane,
highlighting especially the role of the smeared-out domain wall size as an additional length scale 
in the regime of strong quantum fluctuations as well as the anisotropic space-time scaling of the zero-temperature 
critical point, both of which affect the discernability of the thermal KT transition 
at large transverse field. 

The layout is as follows. 
In Sect.~\ref{sec:method}, details of the quantum Monte Carlo method and 
the implementation of periodic boundary conditions are stated. 
In combination with a short review of the well-known KT physics in the inverse square ferromagnetic Ising chain,
Sect.~\ref{sec:BKT_Gamma} provides numerical evidence that the thermal, floating Kosterlitz-Thouless phase 
survives in the presence of a transverse field
and discusses how the smeared-out domain wall size affects the  length scales on which the KT phase can be observed
for large transverse field. Sect.~\ref{sec:summary} gives a summary. 
In Appendix \ref{app:boundstate_size} a variational calculation of the size of a bound state of domain walls (kink and antikink)
is provided. 

\section{Numerical method}
\label{sec:method}

I have used the stochastic series expansion (SSE) quantum Monte Carlo 
method with both single-site quantum cluster updates and multibranch 
cluster updates as described in Ref.~\cite{Sandvik2003} for transverse-field
Ising models with long-range interactions. 
The CPU time scales with the number of spins $N$ as $\mathcal{O}(N \ln N)$
due to an efficient two-step sampling process \cite{Luijten1995, Sandvik2003, Fukui2009}
in which SSE bond operators are sampled from a precomputed discrete probability distribution
and inserted into the SSE operator string depending on the current spin configuration.

For systems with long-range interactions the implementation of boundary conditions 
requires special care. 
Finite-size effets can be minimized through Ewald summation 
which corresponds to replacing the ``bare'' interactions $J_{ij} = 1/(i-j)^{\alpha}$ 
with the sum over all periodic images of the simulation cell 
\begin{equation}
J_{ij}^{(\alpha)} = \sum_{n=-I}^{I} (i - j - nL)^{-\alpha},
\label{eq:PBC_summation_images_numerically}
\end{equation}
where $I$ is chosen large enough for convergence. 
For decay exponent $\alpha=2$ the 
summation over periodic images can be performed analytically 
resulting in the following periodic boundary conditions 
for the inverse square ferromagnetic chain \cite{Fukui2009}
\begin{equation}
 J_{ij} = J \sum_{n=-\infty}^{\infty} \frac{1}{(i-j-nL)^2} = \frac{J}{\zeta^2(|i-j|)},
 \label{eq:PBC_summation_images}
\end{equation}
where the chord length $\zeta(r)$ is defined as 
\begin{equation}
 \zeta(r) \equiv \frac{L}{\pi} \sin\left( \frac{\pi r}{L} \right).
 \label{eq:cord_length}
\end{equation}
In simulations where only one periodic image was used \cite{Sandvik2003}
crossings of the squared magnetization as a function of system size were observed
which were at odds with the expected finite size scaling \cite{Sandvik2003}. 
It has been verified that these crossings are indeed 
due to the choice of boundary conditions (see Appendix ~\ref{app:PBC} for simulations 
with the same PBC as in Ref.~\cite{Sandvik2003}) and disappear if Eq.~\ref{eq:PBC_summation_images}
is used instead. 	
In the following, all energy scales are given in units of $J$.

\section{Line of Kosterlitz-Thouless transitions}
\label{sec:BKT_Gamma}
\begin{figure}[t!]
  \centering
  \includegraphics[width=1.0\linewidth]{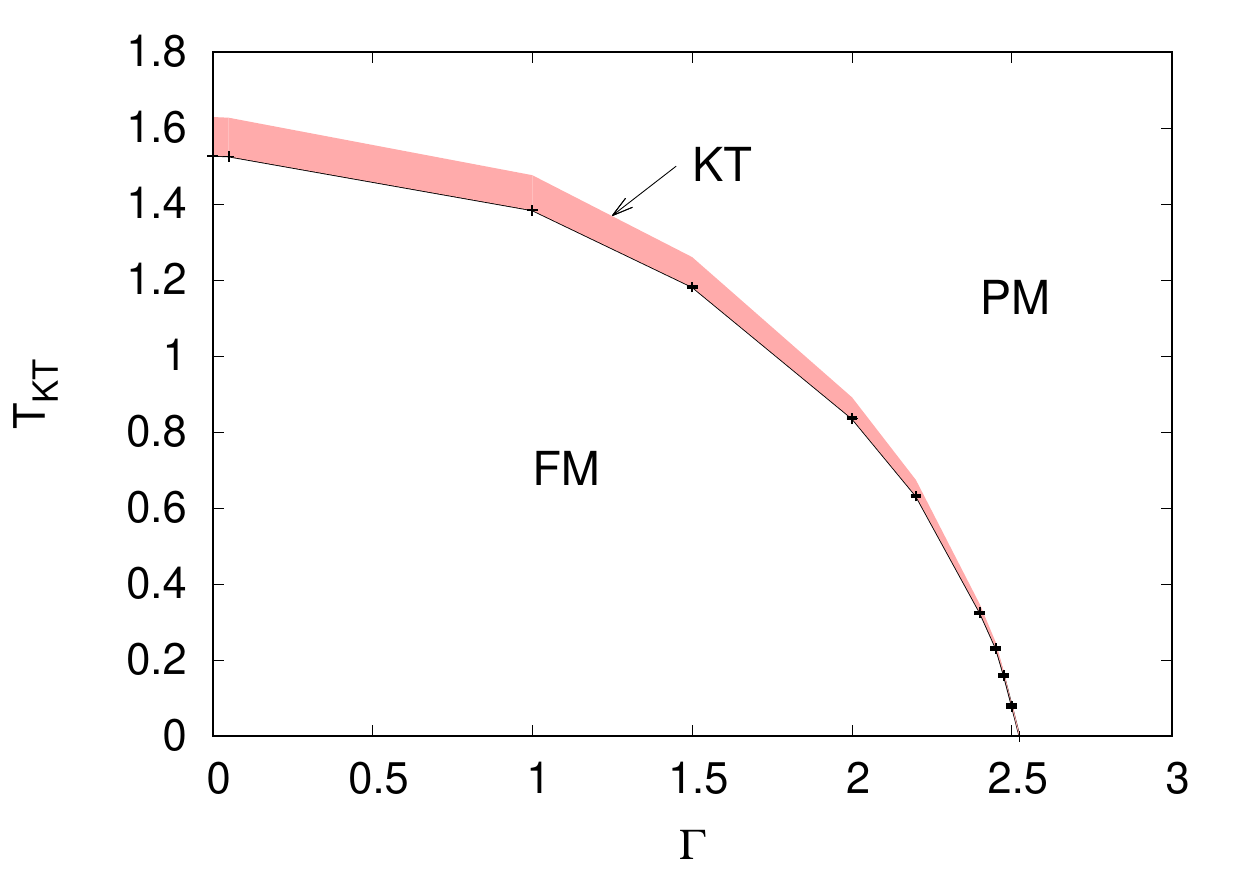}
  \caption[Phase boundary of the inverse-square ferromagnet in a transverse field. ]
  {Phase boundary of the inverse-square ferromagnet in a transverse field.  
  The red area of width  \cite{Bhattacharjee1981, Dutta2001}  $T_{KT1} < T < T_{KT2}^{(RG)} = \frac{16}{15} T_{KT1}$ 
  indicates the floating Kosterlitz-Thouless phase (KT) 
  with infinite susceptibility and algebraically decaying spin-spin
  correlation function \cite{Bhattacharjee1981}; the indicated width of the floating 
  KT phase is a loose upper bound (see main text). 
  Below the red area, there is long-range ferromagnetic order (FM), above the red area the chain 
  is in the paramagnetic phase (PM). At the phase boundary, indicated by the black line,
  there is a universal jump in the magnetization density. 
  The quantum critical point at $\Gamma = 2.524$ \cite{Fukui2009, SyngeTodo_unpublished} is 
  that of an order-disorder transition. Temperature and transverse field are measured in units of $J$.}
  \label{fig:KT_phase_boundary}
\end{figure}

It is well-known that the one-dimensional classical Ising model with 
short-range interactions has no long-range order 
at any finite temperature $T$ \cite{Ruelle1968}
due to competition between the tendency towards alignment to minimize the interaction energy $E$
and the tendency to randomization of the spin configuration to maximize entropy $S$.
In one dimension the tendency towards alignment alway loses in the minimization
of the free energy $F = E - TS$, because there are not enough neighbours. 
As proven by Dyson \cite{Dyson1969}, the balance between energy and 
entropy can be shifted in favour of alignment 
by long-range interactions so that for interactions decaying 
more slowly than $1/r^{\alpha}$ with $\alpha \le 2$, there is true 
long-range order at finite temperature \cite{Ruelle1968}. 
The boundary case $\alpha=2$ is special in that the magnetization
cannot go continuously to zero at the critical temperature \cite{Thouless1969}:
A finite magnetization density $\langle m^2 \rangle = \langle (\frac{1}{N}\sum_i \hat{S}_i^z)^2 \rangle$,
which - in a long-range interaction system - entails an attractive potential 
between domain walls (kinks and antikinks),
is necessary to provide an energy barrier against the entropy-drived 
proliferation of domain walls.
It turns out that at the critical temperature $T_{KT1}$ the ratio 
\begin{equation}
 \frac{\langle m^2 \rangle (T_{KT1})}{T_{KT1}} = \frac{1}{2}
 \label{eq:universal_jump}
\end{equation}
is universal \cite{Luijten2001}, i.e. it does not depend on microscopic details of the model. 
In the present case the universality of the magnetization jump is seen from the 
fact that Eq.~\eqref{eq:universal_jump} holds for any value of the transverse field \cite{Fukui2009},
which indeed does not affect the universality class, as argued in Refs.~\cite{Dutta2001}
and discussed in more detail below.

\subsection{Renormalization group picture}

\begin{figure}
 \centering 
  \includegraphics[width=0.9\linewidth]{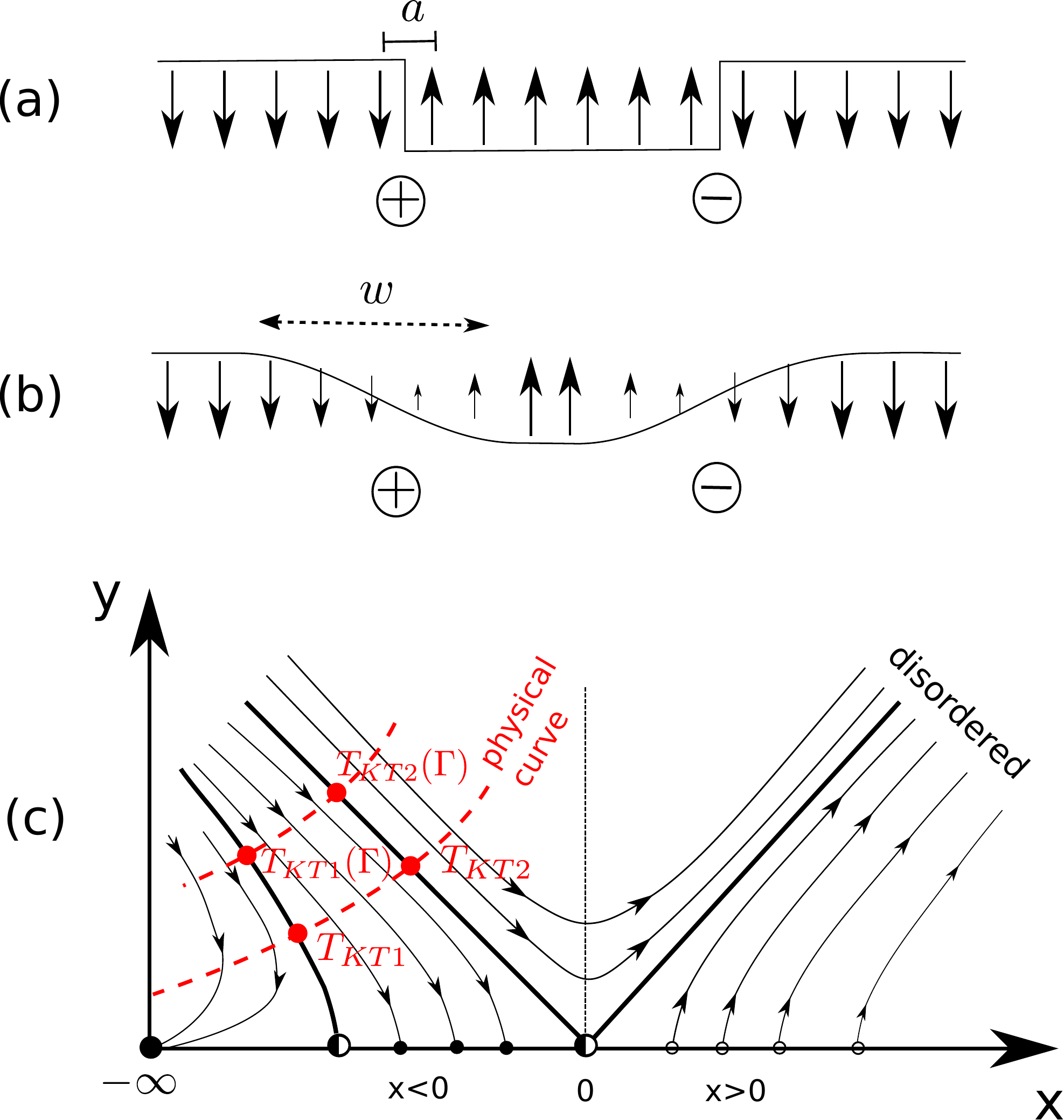}
  \caption[Domain walls in the classical and quantum Ising model.]
  {Domain walls in the classical Ising model (a) and in the transverse-field Ising model (b) with $\Gamma \ne 0$,
  where domain walls are smeared out on a length scale $w$.
  (c) Sketch of the Kosterlitz renormalization group flow adapted to the 
  inverse square quantum ferromagnetic chain.
  $T_{KT1}$ and $T_{KT2}$ ($T_{KT1}(\Gamma)$ and $T_{KT2}(\Gamma)$)
  are the lower and upper critical temperature of the floating KT phase
  in the absence (presence) of a transverse field.}
 \label{fig:kink_antikink}
\end{figure}

The spin configurations of the Ising model can be mapped onto a representation
in terms of domain walls, which is illustrated in Fig.~\ref{fig:kink_antikink}(a-b).
For $1/r^2$ long-range ferromagnetic spin-spin interactions, kinks (denoted as $\textcircled{+}$)
and antikinks (denoted as $\textcircled{-}$), which always appear alternatingly, 
can be regarded as positive and negative electric charges that interact asymptotically via an 
electrostatic potential that depends logarithmically on their distance \cite{Kosterlitz1976, Cardy1981, Bhattacharjee1981}. 
As a consequence of the logarithmic interactions the renormalization group equations 
\cite{Anderson1970, Anderson1971, Kosterlitz1974, Bhattacharjee1981} resemble those 
of the classical XY model, featuring the famous critical KT phase, 
with the difference that for the $1/r^2$ ferromagnet there is true long-range order at finite temperature
which results in a floating critical phase, terminated by two KT transitions 
at temperatures $T_{KT1}$ and $T_{KT2}$. The Kosterlitz renormalization group flow \cite{OrtizBook}
in its adaptation to the $1/r^2$ ferromagnetic chain is sketched in Fig.~\ref{fig:kink_antikink}(c).
There are two scaling variables \cite{Bhattacharjee1981}: the temperature field $x = 1 - \frac{J}{k_B T}$
and the kink fugacity $y = e^{-E_{\text{kink}}/(k_B T)}$, describing the probability for exciting 
a domain wall which involves an energy cost $E_{\text{kink}}$ due to the microscopic deformation 
of the order parameter field (see Fig.~\ref{fig:kink_antikink}(a-b)). 
The characteristic feature of the KT phase is that the kink fugacity is renormalized to 
zero under coarse graining and the trajectories flow to a line of stable fixed points 
on the x-axis for $x<0$, which correspond to a critical phase with infinite susceptibility \cite{OrtizBook}. 
For temperatures above the KT transition $T_{KT2}$ the trajectories flow towards the disordered phase. 
For temperatures below $T_{KT1}$ there is a long-range ordered ferromagnetic phase. 
RG calculations predict a relation of $T_{KT2}^{(RG)}=\frac{16}{15} T_{KT1}$
between the lower and upper transition temperature \cite{Bhattacharjee1981,Dutta2001}.

A transverse field affects the energy  $E_{\text{kink}} = E_{\text{kink}}(\Gamma)$ for the 
local formation of a domain wall, which in the presence of a transverse 
field appears smeared out (Fig.~\ref{fig:kink_antikink}(b)). 
As was argued in Ref.~\cite{Dutta2001}, 
the phase transition remains of the Kosterlitz-Thouless type with critical temperatures 
$T_{KT1}(\Gamma)$ and $T_{KT2}(\Gamma)$ depending on the transverse field $\Gamma$. 
A transverse-field term merely shifts the location of the physical curve
of initial conditions in the RG flow 
through its effect on $E_{\text{kink}}(\Gamma)$ (see upper red dashed line in Fig.~\ref{fig:kink_antikink}(c)),
but it does not change the fix point structure of the RG flow. 
However, more coarse graining steps are necessary to reach the fixed points starting from
the physical curve.

\subsection{Exponentially diverging correlation length}

As the temperature approaches $T_{KT1}$ ($T_{KT2}$) from 
below (above) the correlation length, and also the susceptibility, 
diverge exponentially according to \cite{Kosterlitz1974}
\begin{equation}
 \xi_{T \rightarrow T_{KT1/2}^{\mp}} \sim \exp\left( +\frac{\text{const}}{\sqrt{(|T-T_{KT1/2}|)/T_{KT1/2}}}\right).
 \label{eq:KT_divergence_xi}
\end{equation}
This extremely fast divergence must be contrasted with the power law 
behaviour $\xi \sim t^{-\nu}$ with correlation length exponent $\nu$
of conventional second-order phase transitions. 
For simulations on finite systems, this implies very slow convergence of the 
system size dependent critical temperature $T^{*}(L)$
with linear system size $L$ to the value in the thermodynamic limit 
\begin{equation}
 |T^{*}(L) - T_{KT1/2}| \sim \frac{1}{\log^2(L)},
 \label{eq:KT_finite_size_shift}
\end{equation}
which can be seen from Eq.~\eqref{eq:KT_divergence_xi} by replacing $\xi$ with the maximally attainable 
length $L$ of the finite system. In ordinary finite-size scaling 
there is a more benign power law shift $(T^{*}(L) - T_c) \sim L^{1/\nu}$.

\begin{figure}
\centering 
 \includegraphics[width=1.0\linewidth]{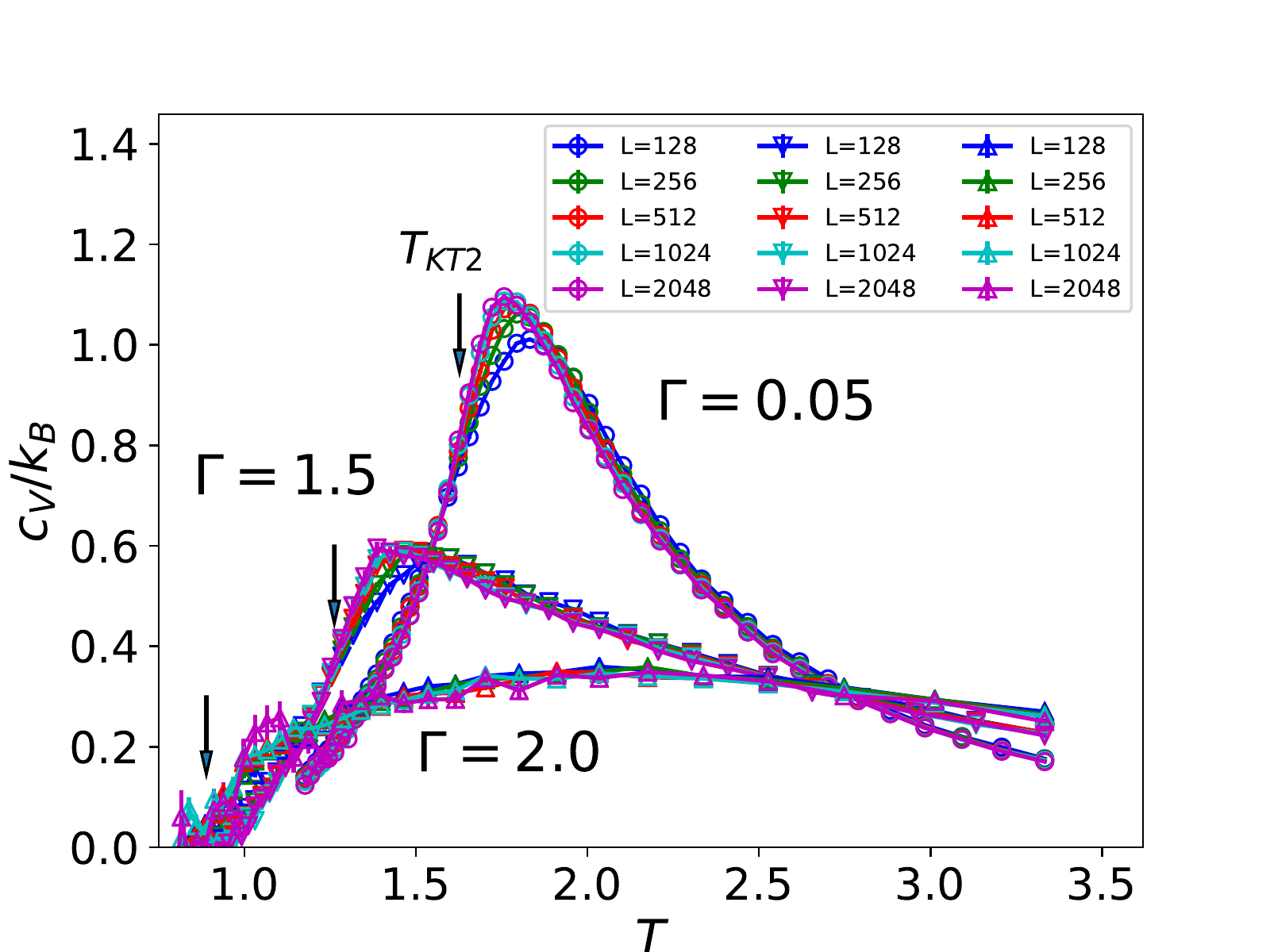}
 \caption[Specific heat in the $1/r^2$ ferromagnetic chain in a transverse field.]
 {Specific heat per site in the $1/r^2$ ferromagnetic chain in a transverse field.
 The arrows show the RG estimate of the upper Kosterlitz-Thouless transition 
 temperatures $T_{KT2}^{(RG)}(\Gamma) = \frac{16}{15}T_{KT1}(\Gamma)$ for the three indicated 
 transverse fields $\Gamma=0.05, 1.5,$ and $2.0$. The non-universal bump
 in the specific heat occurs above the transition temperature and is 
 due to the entropy that is generated when kink-antikink pairs unbind.
 The essential singularity of Eq.~\eqref{eq:KT_singularity_CV} at the transition temperature is not discernible.}
 \label{fig:specific_heat_KT}
\end{figure}
Fig.~\ref{fig:specific_heat_KT} shows the specific heat $C_V= \frac{1}{k_B T^2} (\langle E^2 \rangle - \langle E \rangle ^2)$,
computed from the fluctuations of the energy, 
for different system sizes and transverse fields. 
Similarly to the Kosterlitz-Thouless transition in the 
2D XY model \cite{ChaikinLubenskyBook},
the peak in the specific heat occurs at a higher temperature than $T_{KT2}$, 
the upper transition temperature of the floating KT phase,
and saturates with increasing system size \cite{Bhattacharjee1981}. The renormalization-group prediction 
$T_{KT2}^{(RG)}(\Gamma)$ is indicated in Fig.~\ref{fig:specific_heat_KT} by arrows. 
The singular part of the specific heat, being related to the free energy density $f \sim \xi_{+}^{-1}$
of unbound kinks, vanishes as \cite{ChaikinLubenskyBook}
\begin{equation}
 C_V^{\text{sing}}(T) \sim \xi_{+}^{-1} \sim \exp\left( -\frac{\text{const}}{\sqrt{(T - T_{KT2})/T_{KT2}}}\right)
 \label{eq:KT_singularity_CV}
\end{equation}
and is essentially unobservable if there is another non-universal contribution to the specific heat. 
The non-universal peak in the 
specific heat above $T_{KT2}$ is the result of 
entropy generation due to the gradual unbinding of kink-antikink pairs,
in accordance with the thermodynamic relation $C_V = T \left(\partial S / \partial T \right)_V$.
The fact that the peak becomes smaller and much broader for increasing transverse field 
is a strong sign that the kinks are spatially smeared out so that their 
unbinding does not create much entropy. 
The large kink size makes the system more susceptible 
to finite-size effects, as is also evidenced by the more pronounced crossings 
of the squared magnetization density as a function of $L$  for larger $\Gamma$,
which are shown in Fig.~\ref{fig:crossings} in Appendix \ref{app:PBC}.
The average size of a kink-antikink bound state as a function of transverse field 
is estimated in Appendix ~\ref{app:boundstate_size}.

\begin{figure}
\centering
 \includegraphics[width=0.9\linewidth]{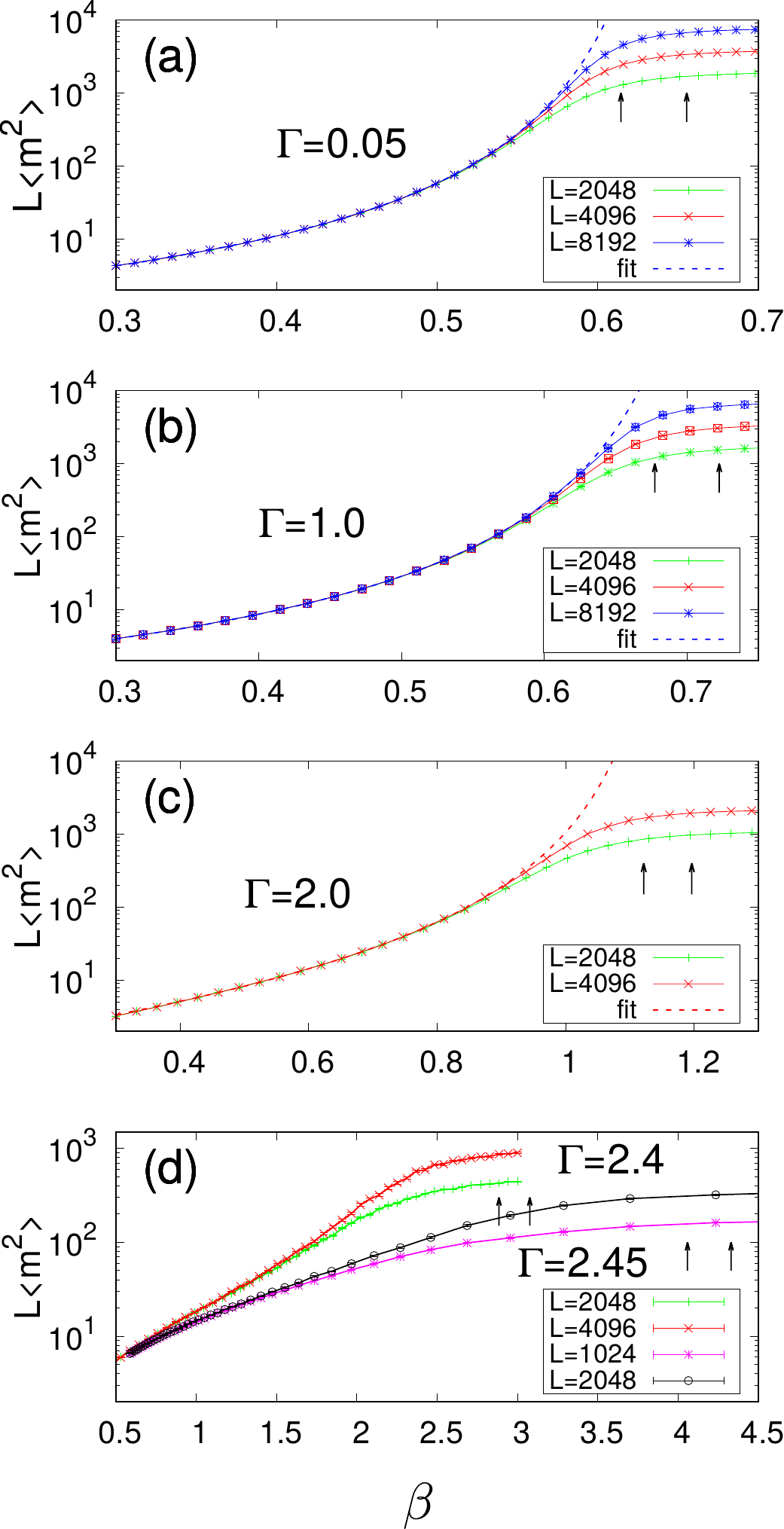} 
 \caption[Exponential divergence of the susceptibility.]
 {Reduced susceptibility $\tilde{\chi}^{(\text{class.})} = L \langle m^2 \rangle$ 
 for different values of the transverse field $\Gamma$.
 Dashed lines are fits to the exponential divergence of Eq.~\eqref{eq:KT_divergence_xi}.
 Arrows indicate $T_{KT1}$ (right) and $T_{KT2}^{(RG)} = \frac{16}{15} T_{KT1}$ (left).}
 \label{fig:suscep_KT}
\end{figure}

The susceptibility is given by the Kubo integral 
\begin{equation}
 \chi^{(\text{qm.})} = L \int_0^{\beta} \langle m(\tau) m(0) \rangle d\tau - L \beta \langle |m| \rangle^2
 \label{eq:suscep_qm}
\end{equation}
where $m=\frac{1}{L} \sum_{i=1}^L S_i^{z}$ is the average magnetization and $m(\tau) = e^{\tau H} m e^{-\tau H}$. 
Since in the transverse field Ising model $[H,m] \ne 0$, this quantity is not identical to 
\begin{equation} 
\chi^{(\text{class.})} =  \beta L(\langle m^2 \rangle - \langle |m| \rangle ^2).
\label{eq:suscep_class}
\end{equation}
However, close to a thermal phase transition the difference $\chi^{(\text{qm.})} - \chi^{(\text{class.})} > 0$  
is a non-diverging quantity and it has been checked on smaller systems that the approximation 
Eq.~\ref{eq:suscep_class} is excellent at elevated temperatures. 
For the purpose of capturing the exponentially diverging 
correlation length the susceptibility was computed
according to Eq.~\eqref{eq:suscep_class}, which is less computationally demanding 
than the Kubo integral over imaginary time in Eq.~\eqref{eq:suscep_qm}. 
Fig.~\ref{fig:suscep_KT} illustrates the exponential divergence with inverse temperature $\beta$
of the reduced susceptibility $\tilde{\chi}^{(\text{class.})} = L \langle m^2 \rangle$, which,
in the disordered phase, scales like 
the square of the correlation length.  Curves are shown, for system sizes up to
$L=8192$ sites, for five different transverse field values $\Gamma$
together with fits to the exponential divergence \cite{Luijten2001} according to Eq.~\eqref{eq:KT_divergence_xi}.
The inverse of the lower and upper KT transition temperatures $T_{KT1}$ (right) and $T_{KT2}^{(RG)} = \frac{16}{15} T_{KT1}$ (left),
which are determined in the following sections, are indicated by arrows.
It was not possible to obtain reliable estimates of $T_{KT2}$ directly from the fit to the exponential 
form in Eq.~\eqref{eq:KT_divergence_xi} due to the limited system sizes (cf. \cite{Luijten2001}).
For $\Gamma=2.4$ and $\Gamma = 2.45$, the exponential divergence is no longer 
clearly visible and the thermal phase transition appears to be masked 
by quantum effects for the given system sizes. 
This suggests that at least up to $\Gamma=2.0$ the Kosterlitz-Thouless scaling relations
which lead to Eq.~\eqref{eq:KT_divergence_xi} hold. 

\subsection{Phase boundary in the T-$\Gamma$ plane}

The shift of the transition temperature $T^{*}(L)$ with system size 
in Eq.~\eqref{eq:KT_finite_size_shift} motivates
the following finite-size scaling form \cite{Weber1988, Harada1997} 
of the universal jump relation Eq.~\eqref{eq:universal_jump}:
\begin{equation}
 \frac{\langle m^2 \rangle (L)}{T_{KT1}(L)} = \frac{1}{2}\left(1 + \frac{A}{\log^2(L/L_0)} \right).
 \label{eq:universal_jump_FSS}
\end{equation}
Here, $L_0$ is some characteristic length of the order of the lattice spacing \cite{Harada1997}.

Standard finite-size scaling, which rests on the algebraic divergence of the correlation length,
cannot be used in the case of a Kosterlitz-Thouless transition, where the correlation length 
diverges exponentially. Instead an alternative scaling form has been suggested 
based on the renormalization group equations \cite{Kosterlitz1974, Weber1988, Harada1997, Fukui2009}:
\begin{equation}
 \frac{2 \langle m^2 \rangle }{T} - 1 = l^{-1} \Psi(t l^2),
 \label{eq:KT_FSS}
\end{equation}
where $\Psi(x)$ is a scaling function, $t= T/T_{KT1} - 1$ is the reduced temperature and $l = \log(L/L_0)$.
We use both Eq.~\eqref{eq:universal_jump_FSS} and a data collapse analysis based on Eq.~\eqref{eq:KT_FSS}
to estimate the Kosterlitz-Thouless transition temperature $T_{KT1}(\Gamma)$ for different transverse field values $\Gamma$
with high accuracy. The phase boundaries are presented in Fig.~\ref{fig:KT_phase_boundary}, which is the main result of this work.
\begin{figure*}[htp]
 \centering
 \includegraphics[width=0.49\textwidth]{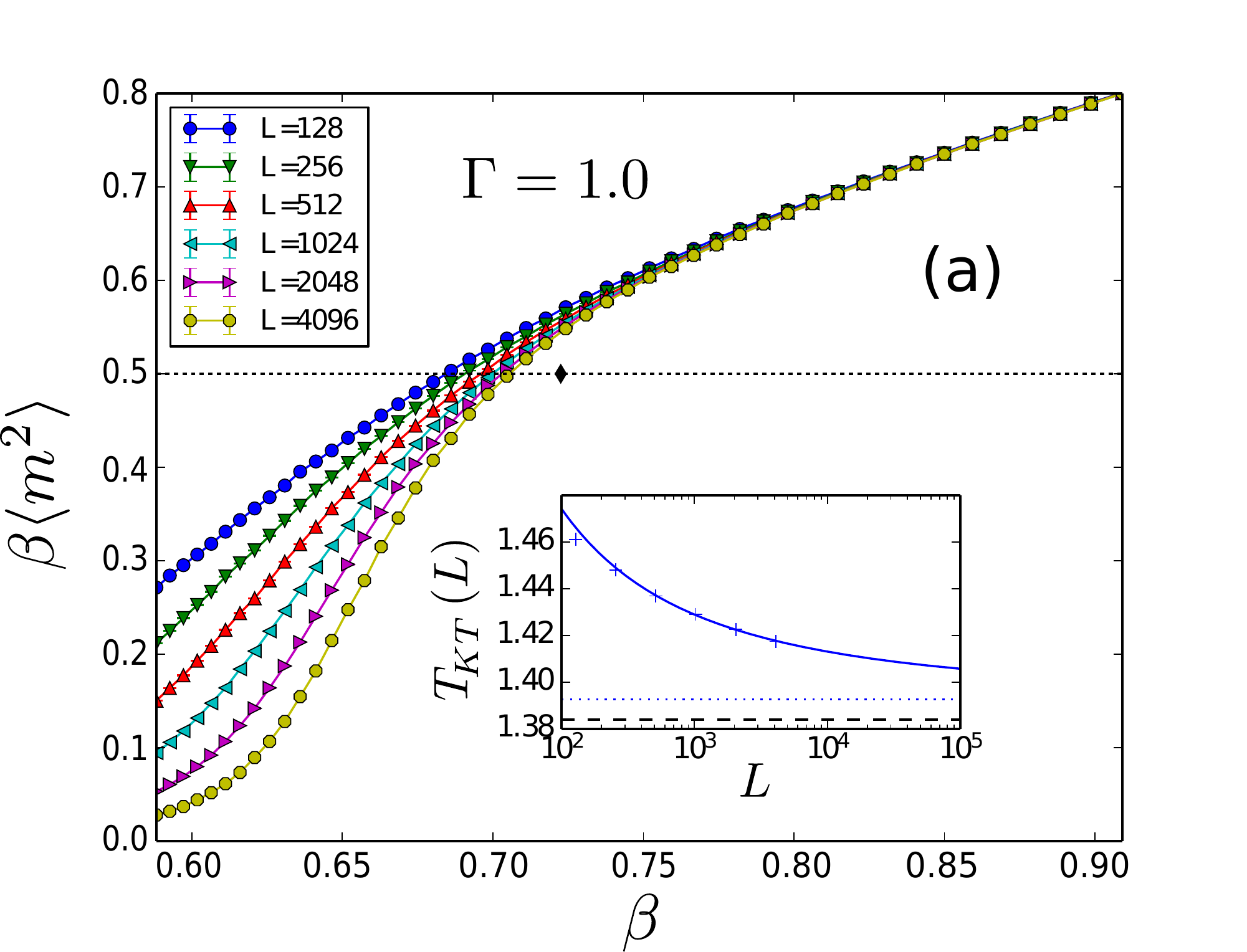} 
 \includegraphics[width=0.49\textwidth]{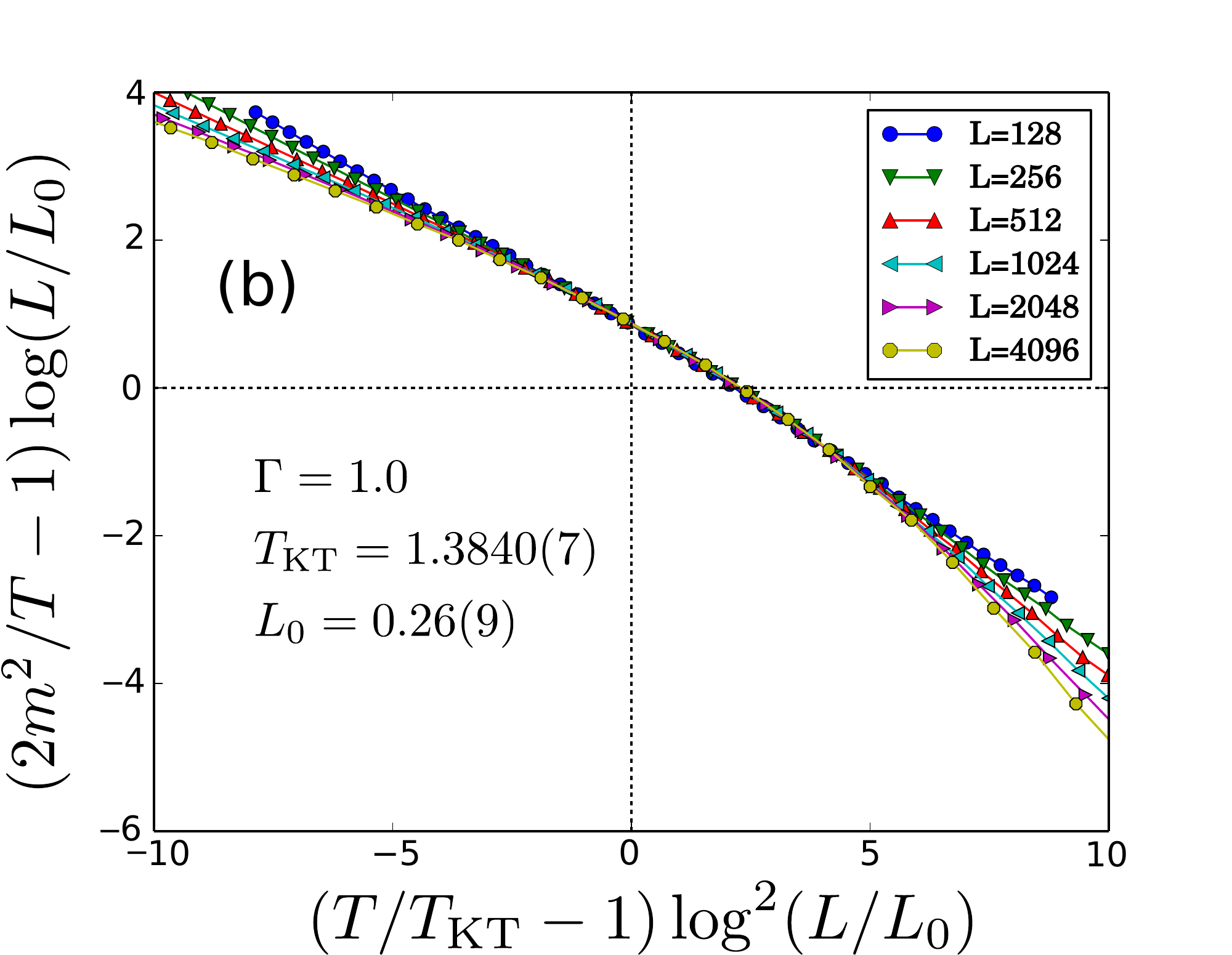} 
 \includegraphics[width=0.49\textwidth]{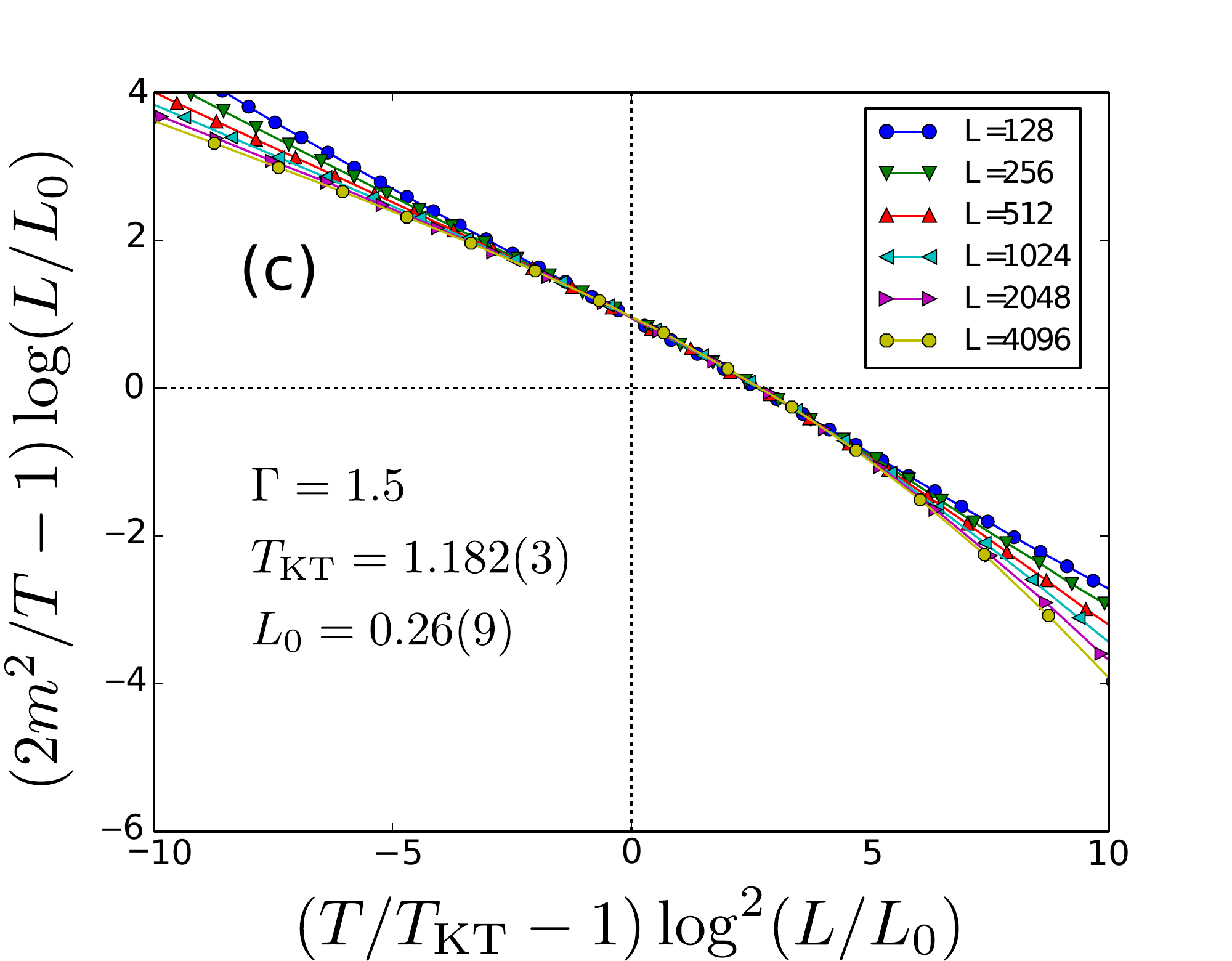} 
 \includegraphics[width=0.49\textwidth]{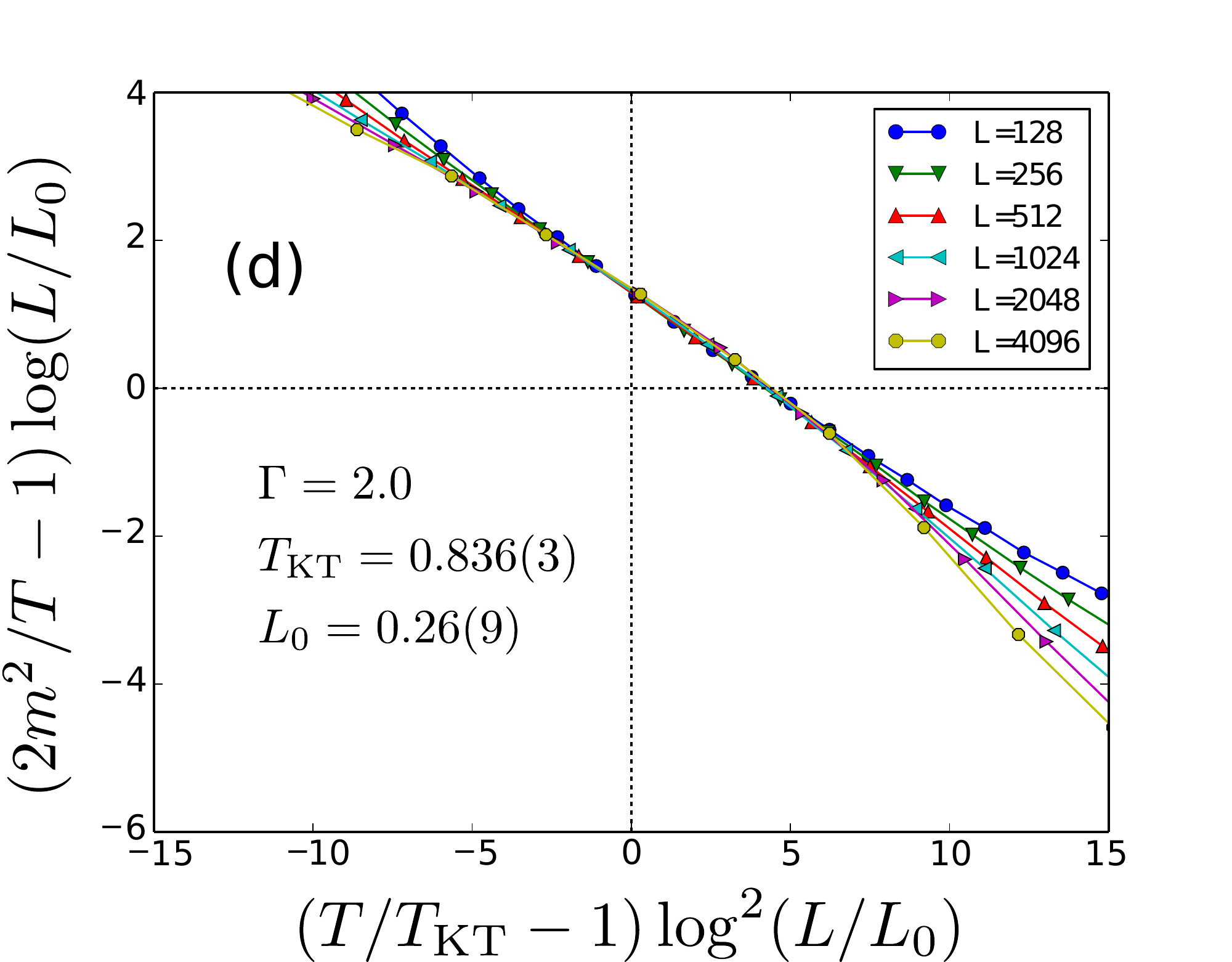} 
 \caption[Data collapse at the Kosterlitz-Thouless transition.]
 {Estimation of the Kosterlitz-Thouless transition temperature $T_{KT} \equiv T_{KT1}$ 
 from an extrapolation of the universal 
 jump relation Eq.~\eqref{eq:universal_jump_FSS} (a) and from a 
 data collapse according to Eq.~\eqref{eq:KT_FSS} (b-d),
 following Ref.~\cite{Fukui2009}.
 The inset in (a) shows a fit of the system-size dependent 
 critical temperature $T_{KT1}(L)$ to $T_{KT1}(L) = T_{KT1}(\infty) + \frac{a}{\log^2(L)}$ for $\Gamma=1.0$.
 (b-d) Scaling plot of the magnetization density for $\Gamma=1.0$ (b), $\Gamma=1.5$ (c), and $\Gamma=2.0$ (d). 
}
 \label{fig:Gamma_KT_collapse}
\end{figure*}
For convenient reference the values of $T_{KT1}(\Gamma)$ and their error bars 
are listed in Tab.~\ref{tab:phase_boundary_KT}.
\begin{table}[t!]
 \centering
 \begin{ruledtabular} 
 \begin{tabular}{@{}l|c|c@{}}
$\Gamma / J$   & $T_{KT1} / J$  & \text{method used or literature Ref.} \\
 0         & 1.52780(9)    & Ref.~\cite{Fukui2009}, see also Ref.~\cite{Luijten2001}   \\
 0.05      & 1.526(4)      & \text{data collapse} \\
 1.0       & 1.3840(7)     & \text{data collapse} \\
 (1.0)     & (1.38460(25)) & (Ref.~\cite{Fukui2009})  \\
 1.5       & 1.182(3)      &  \text{data collapse} \\
 2.0       &  0.836(3)     &  \text{data collapse} \\
 2.2       &  0.632(3)     &  \text{extrapolation} \\
 2.4       &  0.325(3)     &  \text{extrapolation} \\
 2.45      &  0.231(4)     &  \text{extrapolation} \\
 2.475     &  0.160(4)     &  \text{extrapolation} \\
 2.5   	   &  0.079(5)     &  \text{extrapolation} \\
 2.524	   &   0           & quantum-crit. point, Ref.~\cite{Fukui2009, SyngeTodo_unpublished}     \\ 
 \end{tabular}
 \end{ruledtabular}
\caption{Lower boundary $T_{KT1}(\Gamma)$ of the floating KT phase 
as obtained in this work (see Fig.~\ref{fig:KT_phase_boundary})
together with known values from the literature for comparison.
The error bars in brackets indicate the uncertainty in the last digit. }
\label{tab:phase_boundary_KT}
\end{table}

Fig.~\ref{fig:Gamma_KT_collapse}(a) shows the convergence of the magnetization curve
$\langle m^2 \rangle(\beta)$ with system size to the universal jump, $\langle m^2 \rangle = \frac{1}{2} T_{KT1}(\Gamma)$ 
in the thermodynamic limit. The value of the transverse field is $\Gamma = 1.0$.
For comparison,  the diamond indicates the critical
temperature $T_{KT1} = 1.3840(7)$ obtained from a data collapse (see ~\ref{fig:Gamma_KT_collapse}(b))
according to the scaling relation Eq.~\eqref{eq:KT_FSS}. 
When $(2 \langle m^2 \rangle / T - 1) \log(L/L_0)$ is plotted against $(T/T_{KT1} - 1)\log^2(L/L_0)$,
data points for different temperatures $T$
and system sizes $L$ should collapse 
onto a single scaling curve $\Psi(x)$, provided that the critical temperature 
$T_{KT1}$ is chosen correctly. 
Figs.~\ref{fig:Gamma_KT_collapse}(b-d) show the data collapse for $\Gamma = 1.0, 1.5,$ and $2.0$. 
The data collapse is realized with a 
least-squares fit to a polynomical of order eight
with $T_{KT1}$ and $L_0$ as fitting parameters. 
In order to enforce that the fit is particularly good close to the critical point,
a Gaussian weight function is included in the sum of residuals.
The critical temperature $T_{KT1} = 1.3840(7)$ thus obtained for $\Gamma=1$
is in excellent agreement with the only published value at non-zero $\Gamma$, 
$T_{KT1}(\Gamma=1) = 1.38460(25)$ from Ref.~\cite{Fukui2009}, which was obtained with the same type 
of data collapse but including much larger system sizes up to $L = 2^{20} = 1048576$.

The inset in  Fig.~\ref{fig:Gamma_KT_collapse}(a) shows the size-dependent
critical temperature $T_{KT1}(L)$, determined from the universal jump relation Eq.~\eqref{eq:universal_jump}
as the point where $\langle m^2 \rangle (L) / T_{KT1}(L) = \frac{1}{2}$. 
A fit to $T_{KT1}(L) = T_{KT1}(\infty) + \frac{a}{\log^2(L)}$ (see also Ref.~\cite{Harada1998} 
for finite-size scaling of the KT transition in the 2D XY model) verifies that there 
are logarithmic scaling corrections. $T_{KT1}(\infty) = 1.390(3) J$ obtained from this fit is drawn as a dotted line, 
while the dashed line indicates 
the more reliable value for $T_{KT1}$ resulting from the data collapse (see Fig.~\ref{fig:Gamma_KT_collapse}(b)). 
The extrapolation to the thermodynamic limit is inaccurate due to the 
slow, logarithmic convergence with $L$.
For $\Gamma \ge 2.2$ a data collapse was not possible for our system sizes, and
for this range of $\Gamma$ the critical temperatures $T_{KT1}$ presented 
in Fig.~\ref{fig:KT_phase_boundary} are obtained from such extrapolations.

\subsection{Correlation ratio}
For locating the upper KT transition temperature 
the correlation ratio method put forward in \cite{Tomita2002} is used.
By integrating the renormalization group equations it was found 
that in the floating KT phase the connected spin-spin correlation function 
$C_c(r) = \langle S^{z}_r S^{z}_0 \rangle -\langle S^{z}_r\rangle\langle S^{z}_0\rangle$ decays asymptotically according 
to the power law \cite{Bhattacharjee1981}
\begin{equation}
 C_c(r) \sim 4 \frac{\sqrt{|t|}}{r^{4\sqrt{|t|}}} 
 \label{eq:corr_function_asymptotically}
\end{equation}
where $t = (T - T_{KT2}) / T_{KT2} < 0$ is the reduced temperature below the upper KT transition.
Consequently, the ratio $g_c(r_1, r_2)$ of the connected correlation function at two large distances,
e.g. $r_1=L/2$ and $r_2=L/4$, 
\begin{equation}
   g_c(L/2, L/4) = \frac{C_c(r_1=L/2)}{C_c(r_2=L/4)} = \left(\frac{1}{2}\right)^{4 \sqrt{|t|}}
   \label{eq:corr_ratio_asymtotically}
\end{equation}
is independent of system size $L$
and curves of $g_c(L/2,L/4; T)$ for different system sizes collapse onto 
a single curve inside the critical phase $T_{KT1} < T < T_{KT2}$ \cite{Tomita2002}. 
The motivation for using the correlation ratio rather than the 
Binder cumulant \cite{Challa1986} of the structure factor or average magnetization 
is that the former captures only the asymptotic long-distance behaviour of $C_c(r)$ 
whereas the structure factor is a sum over $C(r)$ 
which includes its non-universal behaviour at short distances. 
\begin{figure}
 \includegraphics[width=0.9\linewidth]{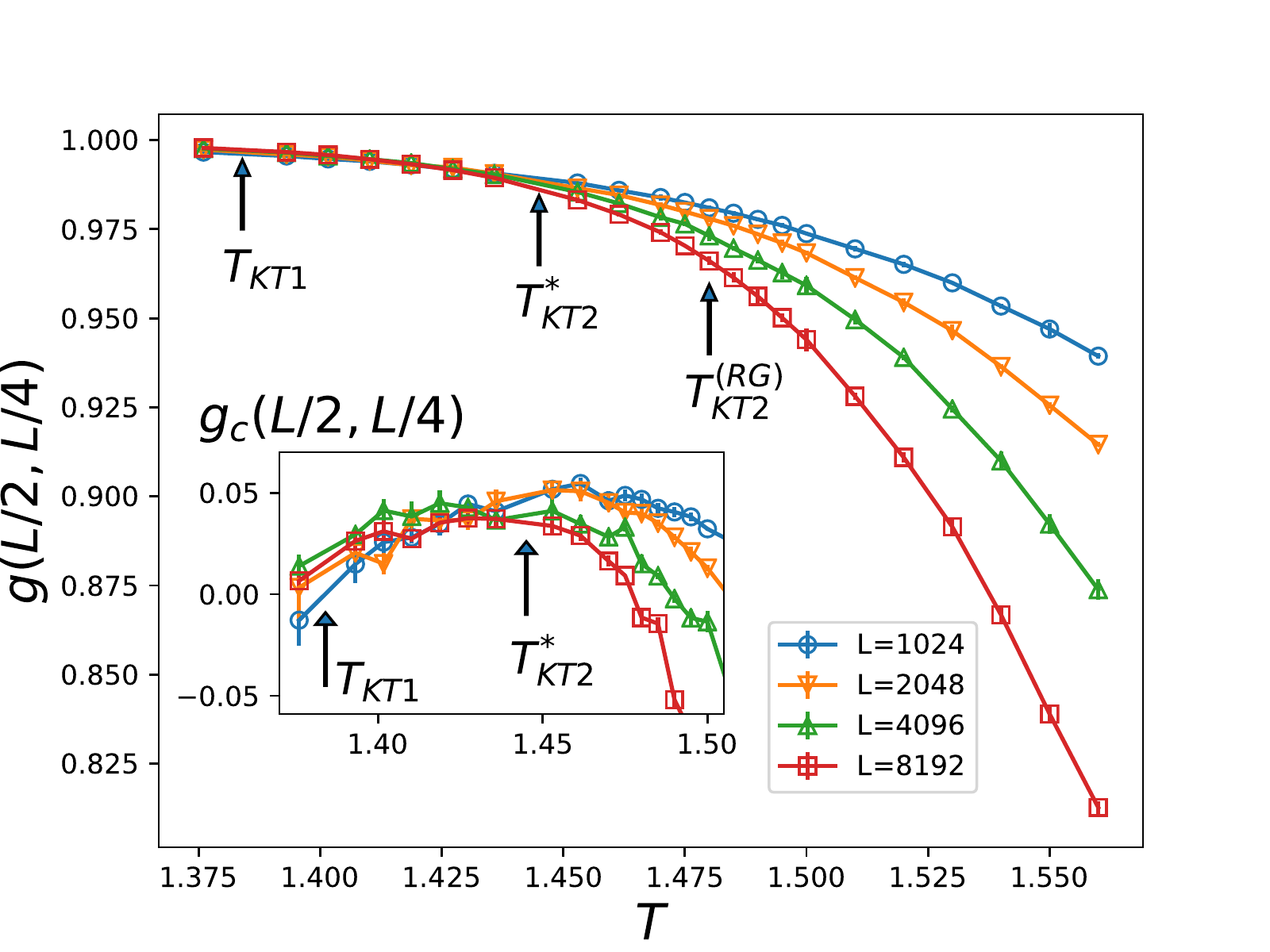}
 \caption{Correlation ratio $g(L/2, L/4)$ as a function of temperature for $\Gamma=1.0$. 
 The arrows indicate $T_{KT1} = 1.3840$, obtained from a data collapse, 
 as well as the renormalization group 
 prediction \cite{Bhattacharjee1981} $T^{(RG)}_{KT2} = \frac{16}{15}T_{KT1}$. The temperature region 
 where the curves start to spray out is marked by $T^{*}_{KT2}$, which is presumably 
 a better estimate of the upper KT transition temperature than $T_{KT2}$.
 The inset shows an enlarged view of the \emph{connected} correlation function $g_c(L/2,L/4)$ in 
 the extended critical region. $T^{*}_{KT2}$ can be defined more precisely as the temperature corresponding 
 to the maximum of the connected correlation function for the largest system size.}
 \label{fig:corr_ratio}
\end{figure}
According to Eq.~\eqref{eq:corr_ratio_asymtotically} the connected 
correlation function $g_c(L/2,L/4)$ should reach a maximum of 1 at $T_{KT2}$, i.e. at $|t| = 0$,
and then decay rapidly as $t$ becomes more negative ($T < T_{KT2}$ but still close to $T_{KT2}$) 
in the extended critical phase. 

Fig.~\ref{fig:corr_ratio} shows the correlation ratio $g(L/2, L/4) = C(L/2) / C(L/4)$ for the ordinary 
correlation function $C(r) = \langle S^{z}_r S^{z}_0 \rangle$ (main panel of Fig.~\ref{fig:corr_ratio})
as well as the connected correlation ratio $g_c(L/2, L/4)$, which was defined 
in Eq.~\eqref{eq:corr_ratio_asymtotically} (inset of Fig.~\ref{fig:corr_ratio}).
As can be seen from the data for $\Gamma=1.0$ in Fig.~\ref{fig:corr_ratio}, 
the correlation ratios for different sizes overlap in
the intermediate KT region and spray out 
for $T > T_{KT2}$. Judging from the main panel of Fig.~\ref{fig:corr_ratio},
the renormalization group prediction \cite{Bhattacharjee1981}
$T_{KT2}(\Gamma=1) = \frac{16}{15} T_{KT1}(\Gamma=1) \approx 1.48$ is a relatively wide upper bound 
for the upper KT transition temperature. 
The collapse of the correlation ratios in Fig.~\ref{fig:corr_ratio} suggests a smaller 
estimate of the upper KT transition temperature as the point 
where the correlation ratios for different system sizes start to fan out, which is approximately indicated 
by the symbol $T_{KT2}^{*}$. If this ``fan-out'' temperature is defined for successive system sizes,
it appears to move to smaller temperature as the system size increases. 
Indeed, based on large-scale Monte Carlo simulations for the classical inverse square 
ferromagnet \cite{Luijten2001} with up to $10^{6}$ sites, 
it has been found that the renormalization group prediction
of Ref.~\cite{Bhattacharjee1981} overestimates the width of the floating KT phase.
The inset of Fig.~\ref{fig:corr_ratio} shows 
that the \emph{connected} correlation function $g_c(L/2, L/4)$ also exhibits a curve collapse, 
with a trend in the temperature dependence that is consistent with Eq.~\eqref{eq:corr_ratio_asymtotically},
namely an increase for \mbox{$t<0$} up to \mbox{$t=0$}, where a maximum is reached
which can be used for the definition of $T^{*}_{KT2} \approx 1.445$.
However, the percise functional form of 
Eq.~\eqref{eq:corr_ratio_asymtotically} is not borne out by the data,
which is probably connected to the fact that Eq.~\eqref{eq:corr_function_asymptotically}
is an asymptotic result which is only valid close to $T_{KT2}$ and 
for system sizes that are orders of magnitude larger (cf. Ref.~\cite{Luijten2001}).
I have not pursued the precise determination of $T^{*}_{KT2}$ for other transverse field values,
and the width of the red margin shown in Fig.~\ref{fig:KT_phase_boundary}
is simply given by the renormalization group prediction $T_{KT2}$, with the understanding that 
it provides a loose upper bound.

\subsection{Influence of the transverse field on the KT transition}
\label{sec:influence_of_TF_on_RG}
It is well established that the thermal phase transition of a quantum system is 
unaffected by quantum fluctuations 
as soon as the diverging correlation length exceeds the length of the \emph{finite} Trotter dimension of the 
effective classical system to which the finite-temperature system is mapped \cite{SachdevBook}; 
hence this length scale drops out. Applied to the present model, it follows directly that the 
KT transition at finite $\Gamma$ should be described by the same  
RG equations as the transition at $\Gamma=0$, with a reduced temperature $T(\Gamma)$ \cite{Dutta2001}.
Indeed, as the phase boundary $T_{KT1}(\Gamma)$ approaches 
the quantum critical point, the universal jump in the magnetization 
decreases and finally vanishes, in agreement with 
the universal jump relation Eq. \eqref{eq:universal_jump}. 
The validity of the KT scaling relations for $\Gamma \le 1$
has been demonstrated numerically in Ref.~\cite{Fukui2009}.
However, close to the quantum critical point the thermal KT phase transition is only observable 
once the correlation length in imaginary time $\xi_{\tau} \sim \xi^z$ exceeds the Trotter dimension,
and due to the space-time anisotropy of the quantum critical 
point with $z=\frac{1}{2}$ \cite{Dutta2001, SyngeTodo_unpublished}, 
this happens only for exceedingly large system sizes
so that, outside a narrow window around the transition temperature, the thermal transition is masked by quantum effects.
In the presence of a transverse field $\Gamma$, a new length scale appears, 
namely the width of a kink $w(\Gamma)$ 
(see Fig.~\ref{fig:kink_antikink}), 
which increases for larger transverse field.
To recover a sharp kink, several coarse graining steps have to be performed
and overall more coarse graining steps are needed to reach scale invariance 
for $\Gamma \ne 0$ than for $\Gamma = 0$. Although a transverse field may change the 
the kink fugacity, i.e. it lowers the microscopic energy cost of inserting a kink,
it will not affect the fixed point structure of the RG flow and the kink fugacity remains an irrelevant 
variable. In combination with the slow divergence of the correlation length 
in imaginary time $\xi_{\tau} \sim \xi^{\frac{1}{2}}$ 
in the vicinity of the anisotropic quantum critical point
and the logarithmic corrections to scaling of Eq.~\eqref{eq:KT_finite_size_shift}, 
this makes the thermal KT transition essentially unobservable for $\Gamma > 2.2$ 
with the system sizes studied in this work.

\section{Summary}
\label{sec:summary}
In conclusion, I have provided precise estimates for the lower transition temperature 
of the floating Kosterlitz-Thouless phase for the $1/r^2$ long-range interacting 
quantum ferromagnet in the entire $T-\Gamma$ plane. 
By considering the ratio of spin-spin correlations at large distances 
the upper transition temperature of the KT phase is shown to be 
significantly smaller than predicted by renormalization group theory, in
agreement with a similar observation in the classical ($\Gamma=0$) system \cite{Luijten2001}.
From the numerical results, at least up to $\Gamma \approx 2$ the nature of the thermal
phase transition can be unambiguously identified as being of the KT type 
while for larger $\Gamma$, down to the quantum critical point, 
the KT scaling is presumably only visible for exceedingly large system sizes and in a tiny 
temperature window, which is a consequence of smeared-out domain walls and 
the strongly anisotropic quantum critical scaling of the correlation lengths in 
space and imaginary time with $z = \frac{1}{2}$.

\section*{Acknowledgments}
The author thanks Yuan Wan for careful reading of the manuscript. 
A large part of this work was done during the author's PhD at 
the University of Stuttgart, Germany. 
The numerical simulations were mostly performed on JURECA, J\"{u}lich Supercomputing Center.
This work is partly supported by the International Young Scientist Fellowship
from Institute of Physics, Chinese Academy of Sciences under the Grant No. 2018004.

\begin{appendix}

\section{Boundary conditions and crossings in $\langle m^{2} \rangle$}\label{app:PBC}
In Ref.~\cite{Sandvik2003} the $1/r^2$ ferromagnet in a transverse field 
was studied with another choice of periodic boundary conditions
\begin{equation}
J_{ij} =  J  \left( \frac{1}{|i-j|^2}  + \frac{1}{(L-|i-j|)^2}\right),
\label{eq:PBC_Sandvik}
\end{equation}
including only one periodic image of the simulation cell,
and concomitantly crossings in the average squared magnetization $\langle m^2 \rangle$ 
were observed as as a function of system length $L$ \cite{Sandvik2003}, which contradicts the finite-size scaling 
that is expected from the model without transverse field \cite{Luijten2001,Sandvik2003}. 
\begin{figure}
 \includegraphics[width=0.8\linewidth]{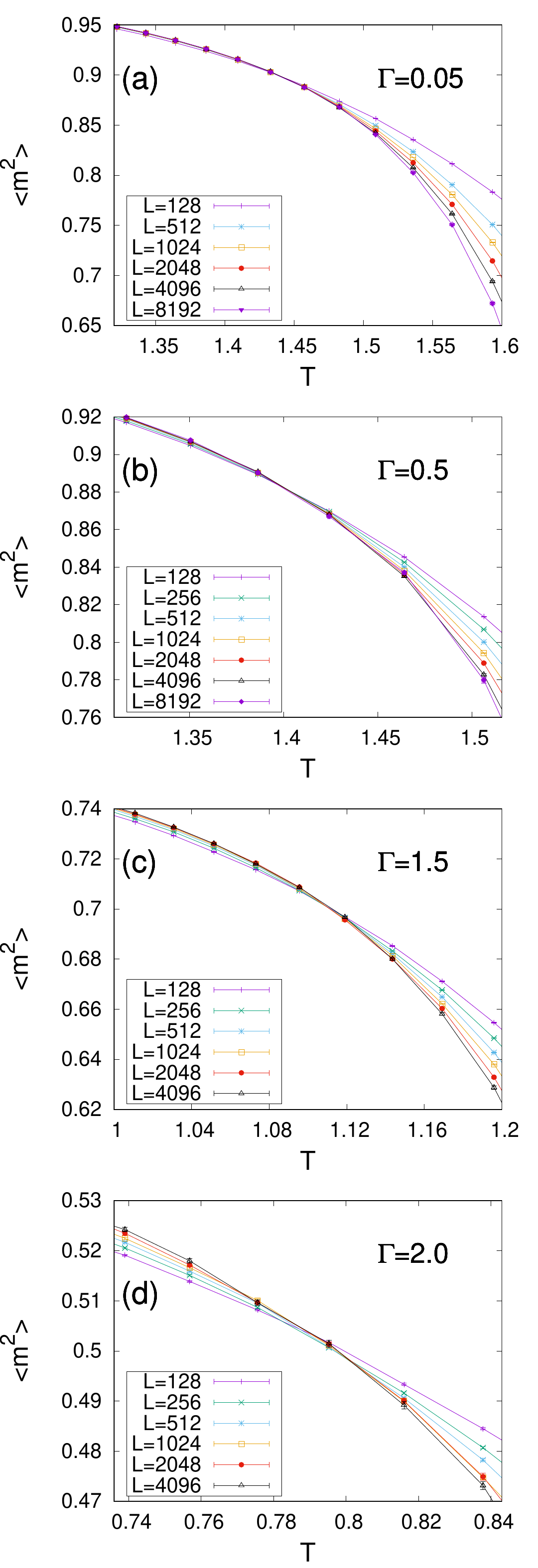}
 \caption{Crossings in $\langle m^2 \rangle$ as a function of system size $L$
 at the Kosterlitz-Thouless transition for $\Gamma=0.05$ (a), $\Gamma=0.5$ (b), $\Gamma=1.5$ (c)
 and $\Gamma=2.0$ (d). Periodic boundary conditions according 
 to Eq.~\eqref{eq:PBC_Sandvik} were used.}
 \label{fig:crossings}
\end{figure}
The hypothesis suggested in 
Ref.~\cite{Fukui2009} that the crossings are due to a relaxation problem in the Monte Carlo calculation of Ref.~\cite{Sandvik2003},
where only a single-site cluster update algorithm was used in the simulations, 
can be ruled out since the same crossings (Fig.~\ref{fig:crossings}) are obtained
with the multi-branch cluster update (also described, but not used for simulations in Ref.~\cite{Sandvik2003})
as long as the periodic boundary conditions of Eq.~\eqref{eq:PBC_Sandvik} are used
rather than those of Eq.~\eqref{eq:PBC_summation_images}. As can be seen from Fig.~\ref{fig:crossings}(a-d)
the crossings become more pronounced as $\Gamma$ increases.
If periodic boundary conditions according to Eq.~\eqref{eq:PBC_summation_images} are used, 
the crossings disappear.

\section{Kink-antikink bound states in the $1/r^2$ ferromagnetic quantum Ising chain}\label{app:boundstate_size}

This appendix provides an estimate of the average size of a bound state of domain walls 
(i.e. a kink bound to an antikink) by means of a variational calculation in a restricted Hilbert 
space.
\begin{figure}[t!]
 \includegraphics[width=1.0\linewidth]{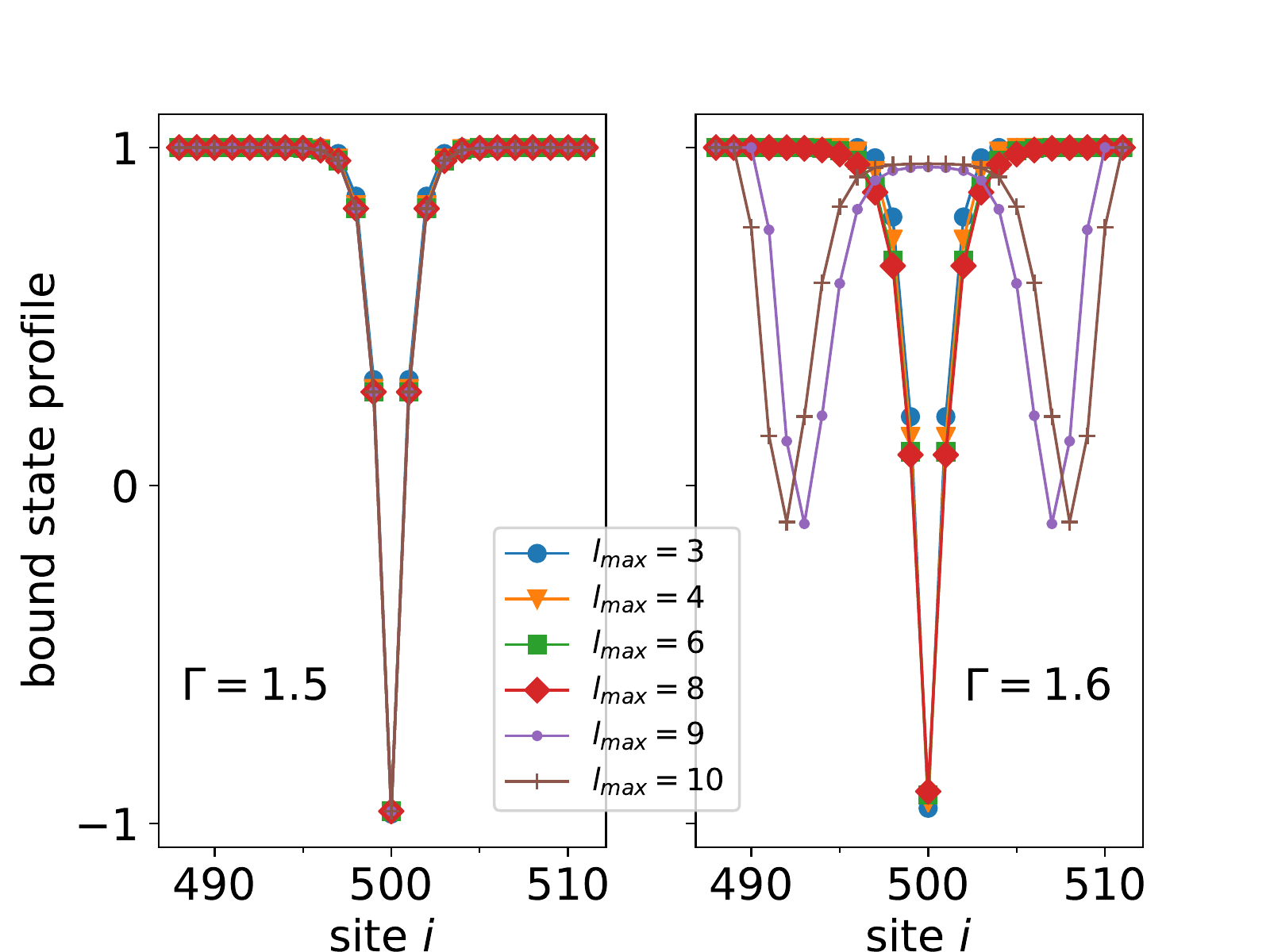}
\caption{Average magnetization profile of a kink-antikink bound state computed from the ground state of 
the Hamiltonian $\tilde{\mathcal{H}}$ projected onto the $n_{\text{max}}=6$
kink subspace for $\Gamma / J_{\text{eff}} = 1.5$ (a) and $\Gamma / J_{\text{eff}}=1.6$ (b).} 
\label{fig:boundstate_profile}
\end{figure}
The subspace containing a single kink-antikink pair is spanned by the basis states
\begin{equation}
 |l, r\rangle = | \uparrow_1 \cdots \uparrow_l \downarrow \downarrow \downarrow_r \uparrow \cdots \uparrow_{L} \rangle.
 \label{eq:basis_state_2kink}
\end{equation}
The length of the chain is set to $L=1001$ and the bound state is forced to be located around the center of the chain $i_c = \frac{L+1}{2}$
by restricting the left and right domain wall positions $l$ and $r$ to 
\begin{equation}
l,r \in \mathcal{I} = [i_c - l_{\text{max}}-1 , i_c + l_{\text{max}} ]
\label{eq:spin_config_2kink}
\end{equation}
with the additional constraint $l < i_c$ and $i_c \le r$.
The action of the transverse-field Ising Hamiltonian $\mathcal{H}$ restricted by the projector $P_{n=2}$ onto the two-kink sector is 
\begin{equation}
\begin{split}
 P_{n=2}\mathcal{H}P_{n=2} |l, r\rangle = -\Gamma ( |l-1, r\rangle + | l+1, r \rangle \\ + | l, r-1 \rangle + |l, r+1 \rangle  ) + V(l,r) |l,r \rangle
\end{split} 
\label{eq:H_2kink}
\end{equation}
Here, the potential energy $V(l,r)$ is given by the interaction energy 
\begin{equation}
 E(\{ S_i^{z} \}) = \sum_{i<j} \frac{J_{\text{eff}}}{\zeta(|j-i|)^2} S_i^{z} S_j^{z}
\end{equation}
of the spin configuration $| \{ S_i^{z}\} \rangle = |l,r\rangle$ according to Eq.~\eqref{eq:basis_state_2kink}, and 
$\zeta(r)$ is the chord length defined in Eq.~\eqref{eq:cord_length}.
Up to a constant energy offset, $V(l,r) \approx A \log(r-l)$ with $A \approx 1.95$ so that the domain walls are confined in the 
two-kink Hilbert space. Here, $J_{\text{eff}} = J \langle m^2 \rangle$, which takes care of the reduction of the magnetization density 
and thus the weakening of the confinement potential due to the presence of kink-antikink bound states at other positions in the chain. 
In the following, energy scales are stated in units of $J_{\text{eff}}$, but the mean-field effect of screened interactions should ultimately be included
by adjusting the energy scales (temperature and transverse field) by a multiplicative factor of $\langle m^2 \rangle$.

For larger transverse field the variational subspace must be extended to include several kink-antikink pairs 
and the Hamiltonian projected onto this subspace of the Hilbert space reads
\begin{equation}
 \tilde{\mathcal{H}} = \sum_{n=2}^{n_{\text{max}}} \tilde{\mathcal{H}}_n + \sum_{n,n^{\prime}}^{n_{\text{max}}} T_{n^{\prime},n},
\end{equation}
where $\tilde{\mathcal{H}}_n = P_n \mathcal{H} P_n$
and $T_{n^{\prime},n} = P_{n^{\prime}} \mathcal{H} P_{n}$
with $P_n$ the projector onto the sector with 
$n=2,4, \ldots, n_{\text{max}}$ kinks. 
The action of the Hamiltonian in the sector of the Hilbert space with $n>2$ domain walls
is analogous to Eq.~\eqref{eq:H_2kink}, with the transverse 
field term inducing the movement of domain walls by one step to the left or to the right and with the constraint 
that the domain wall positions lie in the interval $\mathcal{I}$ defined above.
The transverse-field term also results in off-diagonal matrix elements contained in the matrices $T_{n^{\prime}, n}$ 
between the sectors of $n$ and $n^{\prime} = n \pm 2$ kinks. 
Starting for example from a domain wall configuration $|l,r\rangle$ in the two-kink sector, flipping a spin 
at position $m$ with \mbox{$l+2 \le m \le r-1$} results in a state $|l_1, r_1, l_2, r_2 \rangle$  in the four-kink sector 
with the domain wall positions \mbox{$l_1 < r_1 < l_2 < r_2$} given by $l_1 = l$, $r_1=m-1$, $l_2 = m$, and $r_2 = r$.
Crucially, matrix elements which correspond to creating another domain wall pair outside the bound state spin configuration,
which is defined by the outermost domain walls in a given kink sector, are neglected since otherwise the notion of bound state size becomes ambiguous. 
In the following, $n_{\text{max}} = 6$ is chosen. 

\begin{figure}[h!]
 \includegraphics[width=1.0\linewidth]{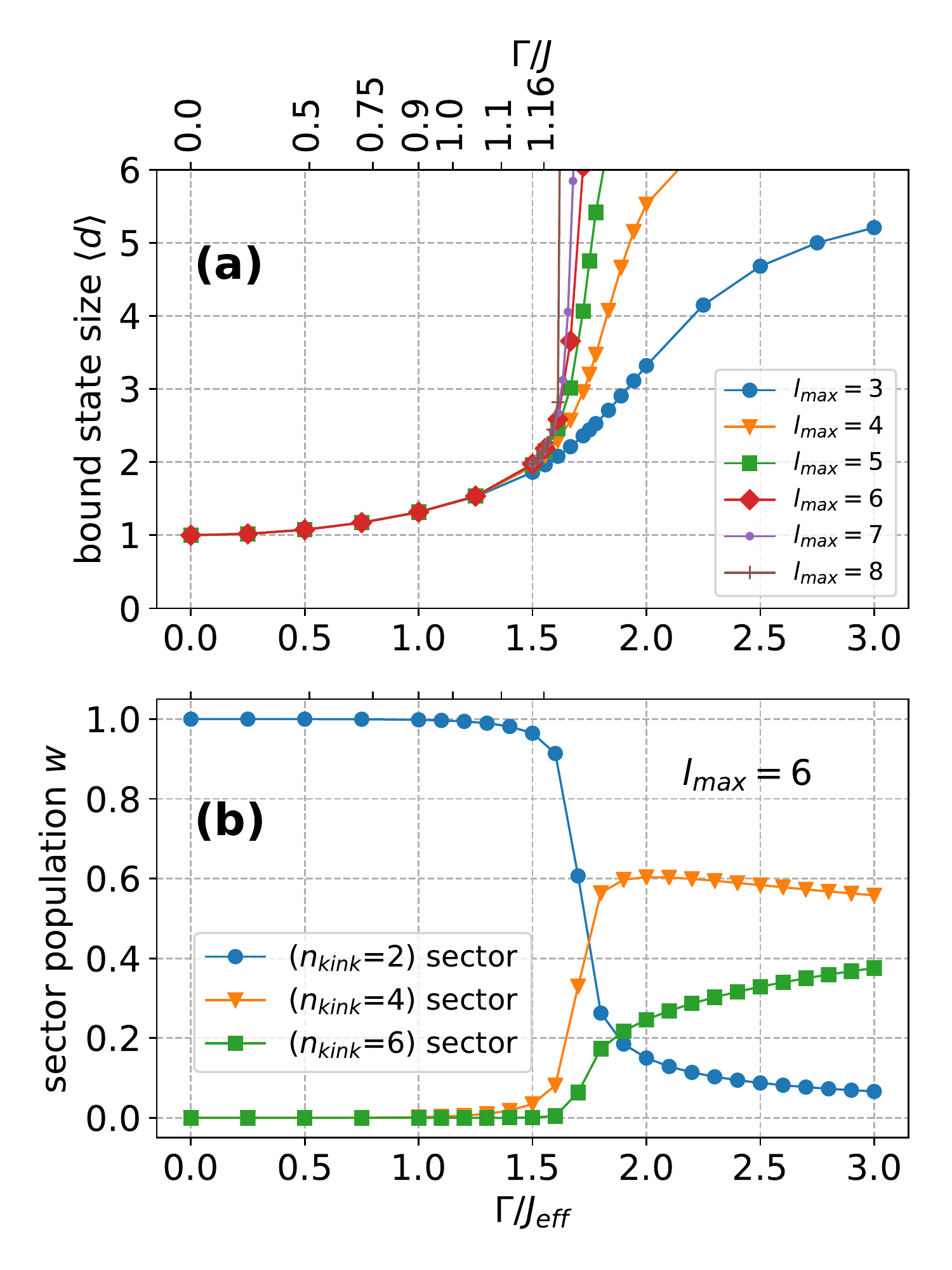}
 \caption{(a) Average size of a kink-antikink bound state, measured as the average distance 
  between the outermost domain walls in the $n_{\text{max}}=6$ subspace (at zero temperature).
 (b) Probability density of the ground state wavefunction in different kink sectors ($l_{\text{max}}=6$).}
 \label{fig:kink_width}
\end{figure}

The spatial extent of a bound state of kink and antikink [Fig.~\ref{fig:kink_width} (a)] is estimated through
\begin{equation}
 \langle d \rangle = \frac{1}{Z} \sum_{m=1}^{M} (r(i)-l(i)) | u(i,m) |^2 e^{-\beta E_m},
\end{equation}
where $Z = \sum_{m=1}^{M} e^{-\beta E_m}$ is the partition sum, $u(i,m)$ 
is the amplitude of the $i$-th basis state in the $m$-th eigenstate of $\tilde{\mathcal{H}}$ with 
eigenenergy $E_m$, and $r(i)$ and $l(i)$ are the positions of the right and left outermost domain walls 
in the $i$-th basis state. 
To assess whether enough kink sectors have been included in the variational subspace, the 
sector populations [Fig.~\ref{fig:kink_width} (b)] are computed as 
\begin{equation}
 w_{\text{n-kink}} = \frac{1}{Z} \sum_{m=1}^{M} \sum_{i \in H_{\text{n-kink}}} | u(i,m) |^2 e^{-\beta E_m},
\end{equation}
where $i \in H_{\text{n-kink}}$ indicates that the $i$-th basis state belongs to the $n$-kink sector 
of the Hilbert space. 

Fig.~\ref{fig:kink_width}(a) shows that the average bound state width in the ground state of $\tilde{\mathcal{H}}$
grows from $\langle d \rangle = 1$ in the classical limit $\Gamma=0$ up to $\langle d \rangle \approx 2$ at $\Gamma / J_{\text{eff}} = 1.5$. 
Up to \mbox{$\Gamma / J_{\text{eff}} = 1.5$} the bound state profile [see left panel of Fig.~\ref{fig:boundstate_profile}]
is converged with respect to enlarging the variational space by increasing $l_{\text{max}}$,
and the support of the wavefunction lies mainly in the two-kink sector [Fig.~\ref{fig:kink_width}(b)].
For $\Gamma / J_{\text{eff}} \ge 1.6$, quantum fluctuations break the bound state into a pair of bound states of width $\langle d \rangle \approx 3-4$ each, 
which move freely as $l_{\text{max}}$ is increased. The breaking of the bound state is mainly due to resonant couplings between different 
kink sectors. For $\Gamma / J_{\text{eff}} > 1.6$, the variational space of the six-kink model employed here is too small [see Fig.~\ref{fig:kink_width}(b)] so that the curves 
in Figs.~\ref{fig:kink_width}(a,b) are not physically meaningful in this parameter region.
To enable comparison with transverse field values given in the main text, the upper
abscissa in Figs.~\ref{fig:kink_width}(a,b) relates the values of $\Gamma / J_{\text{eff}}$ to those 
of $\Gamma / J$ through the implicit relation $\Gamma / J = (\Gamma / J_{\text{eff}}) \times  \langle m^{2} \rangle(\Gamma / J)$,
where the function $\langle m^2 \rangle(\Gamma / J)$ is obtained from a QMC simulation of a chain with $L=1024$ sites at low temperature. 

\end{appendix}



\end{document}